# Surprising pressure-induced magnetic transformations from Helimagnetic order to Antiferromagnetic state in NiI$_2$

Qiye Liu[+,1], Wenjie Su[+,1], Yue Gu[+,2], Xi Zhang[1], Xiuquan Xia[1], Le Wang[1], Ke Xiao[3], Xiaodong Cui[4], Xiaolong Zou[*,5], Bin Xi[*,6], Jia-Wei Mei[*,1], Jun-Feng Dai[*,1,7]

1. Shenzhen Institute for Quantum Science and Engineering, and Department of Physics, Southern University of Science and Technology, Shenzhen, 518055, China
2. State Key Laboratory of Low Dimensional Quantum Physics and Department of Physics, Tsinghua University, Beijing, China
3. NISE Department, Max Planck Institute of Microstructure Physics, Halle, Germany
4. Physics Department, The University of Hong Kong, Pokfulam Road, Hong Kong, China
5. Shenzhen Geim Graphene Center, Tsinghua-Berkeley Shenzhen Institute & Tsinghua Shenzhen International Graduate School, Tsinghua University, Shenzhen 518055, China
6. College of Physics Science and Technology, Yangzhou University, Yangzhou 225002, China
7. Shenzhen Key Laboratory of Quantum Science and Engineering, Shenzhen 518055, China

[+] The authors contribute to this work equally
[*] Corresponding authors: daijf@sustech.edu.cn, meijw@sustech.edu.cn, xibin@yzu.edu.cn, xlzou@sz.tsinghua.edu.cn

**Abstract:**

**Interlayer magnetic interactions play a pivotal role in determining the magnetic arrangement within van der Waals (vdW) magnets, and the remarkable tunability of these interactions through applied pressure further enhances their significance. Here, we investigate NiI$_2$ flakes, a representative vdW magnet, under hydrostatic pressures up to 11 GPa. We reveal a notable increase in magnetic transition temperatures for both helimagnetic and antiferromagnetic states, and find that a reversible transition from helimagnetic to antiferromagnetic (AFM) phases at approximately 7 GPa challenges established theoretical and experimental expectations [1,2]. While the increase in transition temperature aligns with pressure-enhanced overall exchange interaction strengths, we identify the significant role of the second-nearest neighbor interlayer interaction $J_2^\perp$, which competes with intra-layer frustration and favors the AFM state as demonstrated in the Monte Carlo simulations. Experimental and simulated results converge on the existence of an intermediate helimagnetic ordered state in NiI$_2$ before transitioning to the AFM state. These findings underscore the pivotal role of interlayer interactions in shaping the magnetic ground state, providing fresh perspectives for innovative applications in nanoscale magnetic device design.**

**Keywords**

2D ferromagnetic materials, exchange interaction, helimagnetic order, high-pressure second harmonic generation spectroscopy.

**Introduction**

In the domains of condensed matter physics and materials science, van der Waals (vdW) magnets have emerged as an intellectually captivating frontier, revealing a rich tapestry of magnetic phenomena that has ignited profound interest and exploration[3-7]. The complexity and richness of magnetic behaviors within vdW magnets become increasingly prominent as one delves into the intricate interplay among adjacent layers, characterized by a diverse array of interlayer couplings[8-11]. These interlayer couplings, spanning from van der Waals to magnetic interactions, not only govern the magnetic ground state but also bestow upon these materials an extraordinary degree of tunability. This inherent versatility renders vdW magnets multifaceted and holds the tantalizing potential for groundbreaking applications. The precision control of interlayer coupling is attainable through diverse strategies, encompassing the manipulation of stacking order[12,13], gate tuning[14-16], and the application of pressure[17,18].

Transition metal dihalides, exemplified by $NiI_2$, have emerged as subjects of keen interest, owing to their intricate magnetic behaviors and the potential manifestation of multiferroic properties[19]. In the single-layer configuration, each Ni atom forms an octahedral $NiI_6^{4-}$ structure by bonding with six I atoms. Three such single-layers are stacked along the c-axis, resulting in a $CdCl_2$-type layered structure with a centrosymmetric space group[20]. Specific heat capacity and magnetic susceptibility measurements (Figure 1a) on $NiI_2$ single crystals reveal two distinct phase transitions during the cooling process, occurring at approximately 73 K ($T_{N1}$) and 60 K ($T_{N2}$)[21]. At $T_{N1}$, $NiI_2$ undergoes a transition from a trigonal to a monoclinic structure, exhibiting a colinear antiferromagnetic order with a centrosymmetric R3m space group[22]. In the vicinity of $T_{N2}$, a spiral antiferromagnetic order emerges, with the spins rotating within the plane perpendicular to the propagation vector q~ (0.138, 0, 1.457), slanted at a 45-degree angle from the c-axis (left figure in Figure 1b)[23]. Consequently, the helimagnetic state induces spontaneous electric polarization due to the inverse Dzyaloshinskii–Moriya (DM) interaction, as validated by polarization current measurements under an in-plane magnetic field[24]. Consequently, $NiI_2$ has been postulated to possess a multiferroic ground state, primarily attributed to its robust magnetoelectric (ME) effect. Recent debates have revolved around whether this multiferroic state can persist in $NiI_2$ down to the monolayer limit[25-28].

The presence of a helical spin configuration in non-chiral centrosymmetric crystals alludes to the complex exchange interactions among adjacent magnetic atoms in $NiI_2$, stemming primarily from inherent geometric frustration and the presence of a frustrated Kitaev interaction[29]. The ease of tunability between various magnetic states in $NiI_2$ is expected to be governed by the inter-layer couplings (Figure 1b). To this end, we subjected $NiI_2$ flakes, serving as representative vdW magnets, to hydrostatic pressures of up to 11 GPa. This approach enabled us to embark on a comprehensive investigation into the intricate role of van der Waals interlayer magnetic interactions, furthering our understanding of the competition among different alignments of magnetic spins. Under these meticulously controlled pressure conditions, we observed a noteworthy increase in the magnetic transition temperatures for both helimagnetic and antiferromagnetic states under low pressures. Notably, a striking and reversible transition from a helimagnetic to an antiferromagnetic state was noted at approximately 7 GPa. Our Monte Carlo simulations and reported Density Functional Theory calculations offered valuable insights, revealing the rapid strengthening of interlayer interactions under pressure. This phenomenon led to the enhancement of transition temperatures and, consequently, the observed transition from a helimagnetic to an antiferromagnetic state. These findings underscore the pivotal role played by interlayer interactions in shaping the magnetic ground state.

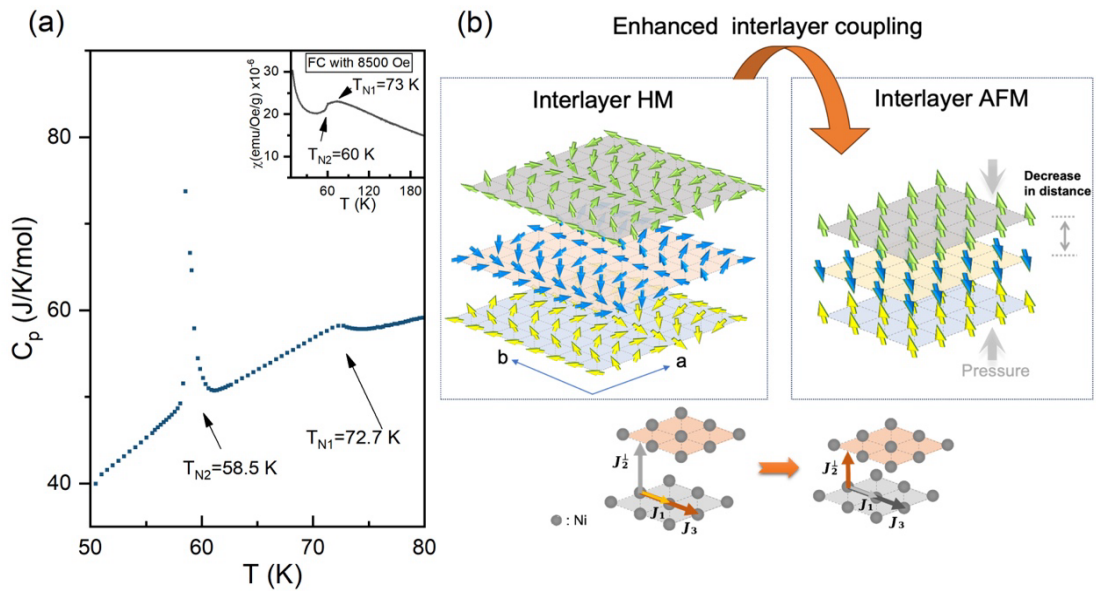

Figure 1 (a) Specific heat capacity and magnetic susceptibility of $NiI_2$ as a function of temperature. FC stands for field cooling. (b) Schematic illustration of the transition of magnetic structure in $NiI_2$, shifting from helimagnetic (HM) to antiferromagnetic (AFM) states due to enhanced interlayer coupling. Lower figure represents the dominate interlayer exchange interactions $J_1$ and $J_3$ at ambient pressure and the perpendicular effective exchange coupling $J_2^{\perp}$ at high pressure.

**Enhanced magnetic transition temperatures under low pressures**

The hydrostatic pressure was applied to the samples using a diamond-anvil cell (DAC) technique, which allowed us to achieve pressures of up to 20 GPa. Initially, $NiI_2$ flakes (50-100 nm) were mechanically exfoliated and then delicately transferred onto the culet of the diamond using a dry transfer method[30]. Subsequently, we introduced a pressure-transmitting medium, namely silicone oil, into the gasket blanks. The DAC was assembled by aligning and securing the two diamonds with four screws. Notably, this entire procedure was conducted within a nitrogen-filled glove box to prevent any degradation of the samples.

Temperature-dependent Raman measurements and polarization-sensitive second harmonic generation (SHG) were carried out in a back-reflected geometry under applied pressures, capitalizing on the excellent transparency and low polarization response of diamond. These techniques were significant in determining the magnetic transition temperatures. The Raman measurements allowed us to extract the magnon intensity evolution as temperatures varied and ascertain the transition temperature of collinear antiferromagnetic orders. Additionally, polarization-resolved and temperature-dependent SHG spectra were obtained to probe the inversion-symmetry-breaking features of the spin structure under pressure, yielding insights into the transition temperature to the helimagnetic state. (For detailed information regarding the experimental setup, please refer to the "Methods" section and Figure S1.)

We conducted temperature-dependent Raman measurements on a $NiI_2$ flake, utilizing an excitation centered at 1.96 eV. Under ambient pressure, six distinct Raman peaks were observed at 15 K (Figure S2), in line with previously reported findings in the literature[27]. Of particular interest is the magnon peak in the vicinity of 120 cm$^{-1}$, where its intensity displays a significant temperature dependence. Specifically, the intensity gradually reduces as the temperature increases, ultimately vanishing between the Néel temperature transition points ($T_{N1}$ and $T_{N2}$). This provides a reliable means to estimate the Néel temperature, despite a slightly lower temperature reading compared to values obtained through thermodynamic measurements.

Figure 2a-2f provides a visual representation of our temperature-dependent Raman results under various hydrostatic pressures. At 15 K, two phonon modes ($P_1$ and $P_2$) and a magnon mode (M) exhibit a linear blueshift as pressure increases (Figure 2g and Figure S3), an effect attributed to enhanced interlayer coupling. Importantly, the absence of new Raman peaks indicates the structural stability of the crystal below 13 GPa. However, it is worth noting that as reported in references [31-33], the structure of $NiI_2$ begins to collapse when the pressure exceeds 19 GPa. The temperature-dependent integrated intensity of the magnon mode under varying pressure conditions is plotted in Figure 2h. Notably, our observations reveal the remarkable resilience of the magnon signals even under elevated pressures and increased temperatures. For instance, at 2.7 GPa, the magnon peak persists at 100 K, and it continues to endure at 115 K under 4.4 GPa, and at 135 K under 6.0 GPa. Surprisingly, at even higher pressures, such as 7 and 8.6 GPa, the magnon signals remain discernible, with their transition temperature showing a remarkable increase, reaching 155 K at 7 GPa and 200 K at 8.6 GPa. It underlines the robustness of the antiferromagnetic state, illustrating its resilience against the influence of pressure. This pressure-induced enhancement of transition temperature has also been reported in other vdW magnets, such as $CrSi(Ge)Te_3$[34,35].

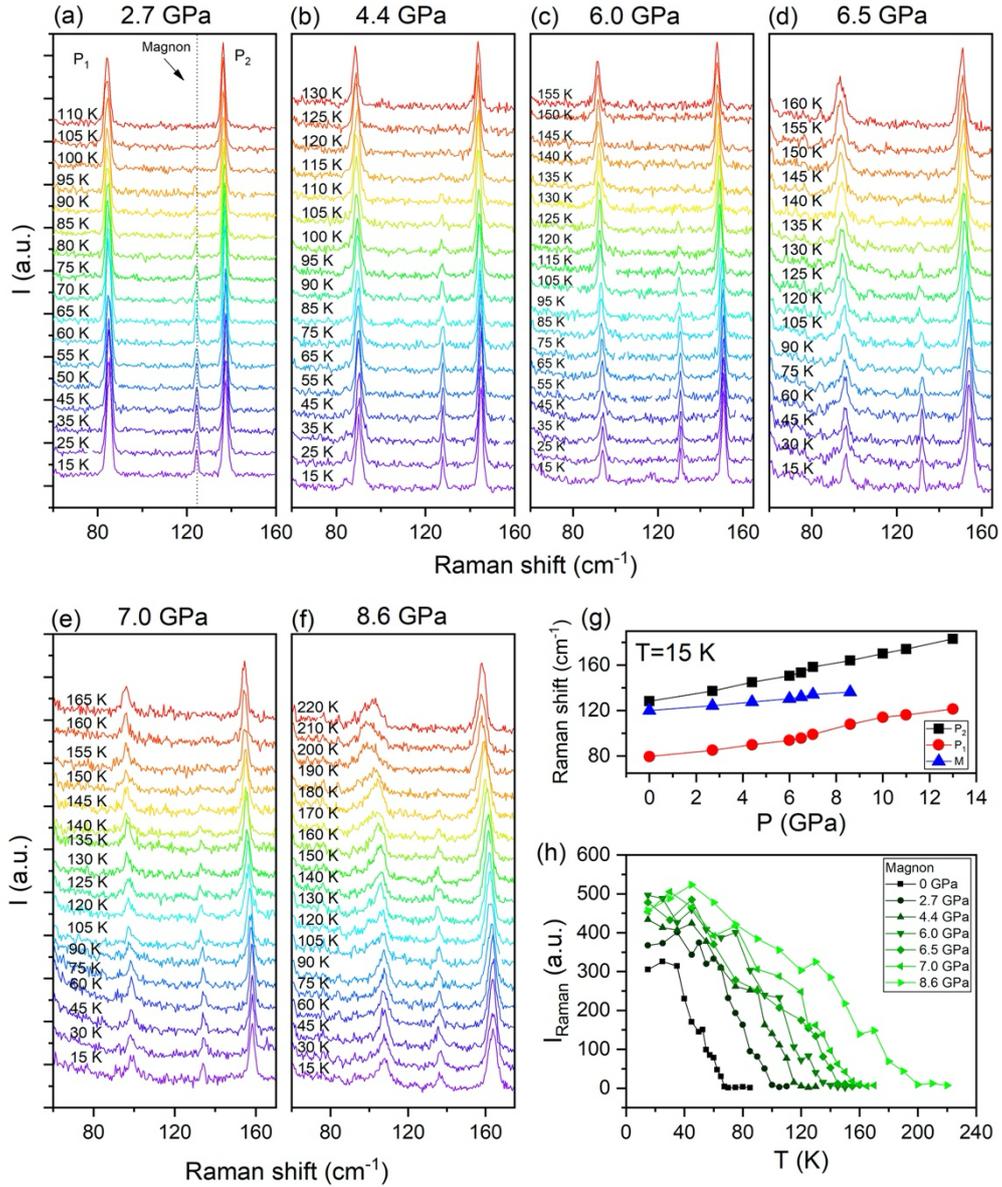

Figure 2 (a)-(f) Raman spectra of NiI$_2$ flake measured at different temperatures under six fixed pressures, e.g., 2.7, 4.4, 6.0, 6.5, 7.0, and 8.6 GPa. The dotted line in (a) indicates the peak of the magnon mode. (g) Raman shifts of two phonons modes and a magnon mode as a function of pressure. (h) The integrated intensity of magnon peak as a function of temperature at 0, 2.7, 4.4, 6.0, 6.5, 7.0, and 8.6 GPa, respectively.

At room temperature and normal pressure, NiI$_2$, characterized by a centrosymmetric space group, inherently exhibits an electric-dipole forbidden second harmonic generation (SHG). However, below the Néel temperature T$_{N2}$, a phase transition occurs, leading to the emergence of helimagnetic orders. This transition signifies the breaking of both spatial inversion and time-reversal symmetries, thereby enabling SHG induced by magnetism. We observe a distinct and robust SHG signal at temperatures below 59 K (Figure S4a), reaching magnitudes up to two orders of magnitude greater than those observed at elevated temperatures. This pronounced SHG signal indicates the presence of a helical spin structure beneath the transition temperature, a characteristic that is consistent with the transition temperature $T_{N2}$ determined by the magnetic susceptibility measurement. Moreover, the polarization-dependent SHG measurement shows that the SHG intensity at 15 K reveals a strong signal in the XY component, manifesting as a distorted "8" pattern, while the XX component exhibits a relatively weaker signal characterized by a distorted butterfly pattern. Here, the XX and XY components indicate that the orientation of the SHG signal is parallel and perpendicular to the polarization of the incident light, respectively. Both SHG patterns unveil a two-fold symmetry in $NiI_2$ induced by the contribution of both electric and magnetic dipoles below $T_{N2}$ (Figure S4b), indicating a mirror symmetry breaking. Furthermore, the polarization-dependent SHG at 65 K reveals a distinctive six-fold rotational symmetry under high-power excitation (Figure S3b). This symmetry reflects a C3 symmetry, signifying that $NiI_2$ has transitioned to a collinear antiferromagnetic order characterized by spatial inversion symmetry. It may originate from the quadrupole term and surface electric dipole term. Consequently, SHG measurements serve as a valuable tool for delineating the phase transition from helimagnetic to collinear antiferromagnetic states.

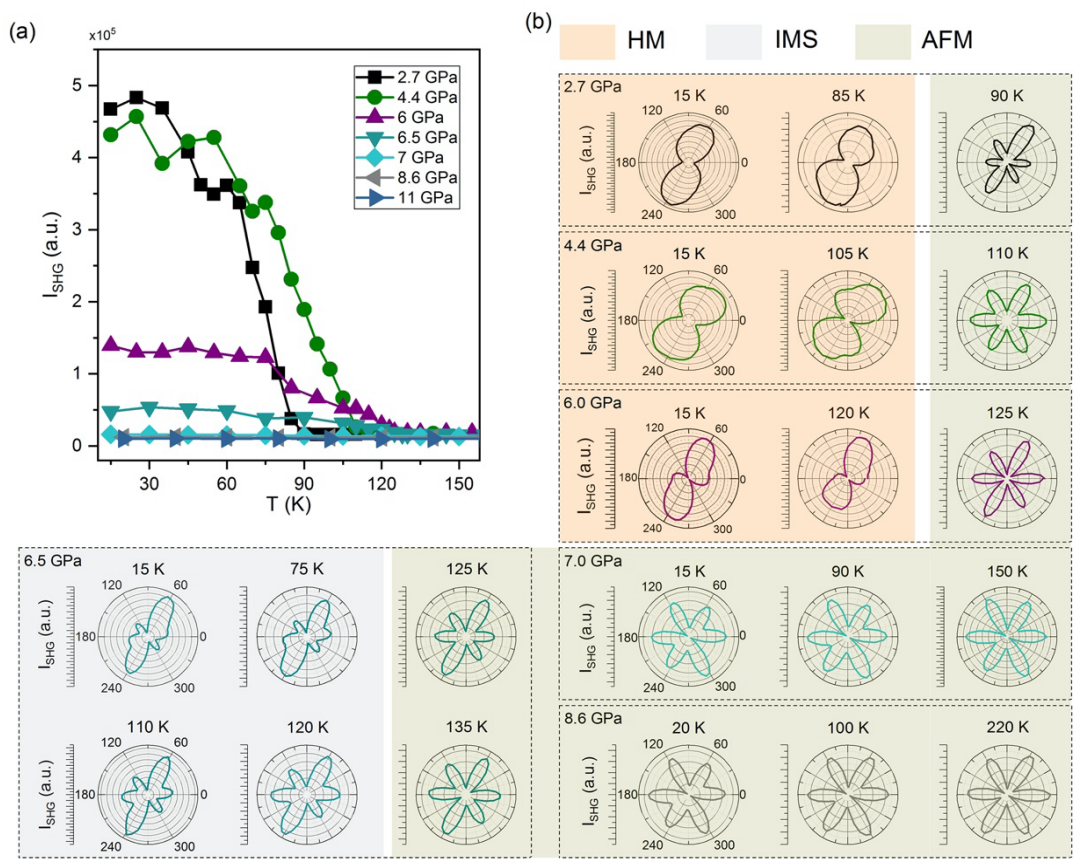

Figure 3 (a) Temperature-dependent SHG intensity in a $NiI_2$ flake at seven fixed hydrostatic pressures, e.g., 2.7, 4.4, 6.0, 6.5, 7.0, 8.6, and 11.0 GPa. (b)-(g) Angular-resolved SHG patterns for temperatures below and above the helimagnetic transition temperature at the six representative hydrostatic pressures. Here, HM, AFM and IMS represent the helimagnetic state, antiferromagnetic state and intermediate state, respectively.

Figure 3a presents the SHG signal intensity as a function of temperature under seven constant hydrostatic pressures. At 2.7 GPa, the SHG signal intensity undergoes a gradual change as the temperature increases from 15 to 65 K. However, it experiences a rapid and significant decrease, approaching a minimum at approximately 90 K. This shift corresponds to a remarkable alteration, with a maximum-to-minimum intensity ratio of approximately 30:1 during the temperature rise. Consequently, we can roughly identify this point as the transition temperature marking the phase transition from the helimagnetic to antiferromagnetic states under pressure. Specifically, at 2.7 GPa, this transition temperature is found to be around 90 K, surpassing the 59 K recorded under normal pressure. Elevating the pressure to 4.4 and 6.0 reveals a similar temperature-dependent SHG signal curve, accompanied by a corresponding increase in the transition temperature (as shown in Figure 3a). The maximum transition temperature is determined to be 125 K at 6 GPa.

Beyond the SHG signal intensity, the SHG pattern also serves as a distinctive indicator for distinguishing between helimagnetic and antiferromagnetic states. To clarify the symmetric information, we only focus on the XY component in the polarization-dependent SHG measurement. As displayed in Figure 3b, the XY component of the polarization-resolved SHG spectrum retains a distorted "8" pattern at both 15 K and 85 K under 2.7 GPa pressure. This signifies a two-fold symmetry that persists at lower pressures. However, at 90 K, the SHG pattern undergoes a transformation, displaying a distorted six-petal pattern indicative of C3 symmetry. It's worth noting that the distortion may be influenced by the weak polarization response of the diamond under pressure. Thus, the transition temperature ($T_{N2}$) is determined to be 87.5±2.5 K for $NiI_2$ under 2.7 GPa pressure, consistent with the value derived from the intensity-temperature relationship. At both 4.4 and 6 GPa in Figure 2b, a noticeable difference in the SHG pattern remains evident as the temperature increases. Consequently, the transition temperature is approximated to be 107.5±2.5 K at 4.4 GPa and 122.5±2.5 K at 6 GPa, allowing for the identification of the maximum transition temperature from the helimagnetic to antiferromagnetic states via the SHG patterns.

**Pressure-induced magnetic phase transition from HM to AFM**
Remarkably, as we escalate the pressure to 7 GPa, an intriguing observation emerges: the SHG intensity exhibits negligible temperature dependence within the wide range of 15-150 K (as indicated by the light green diamonds in Figure 3a). This intriguing temperature independence is sustained even when the pressure is further raised to 8.6 and 11 GPa. The absence of a discernible turning point in this scenario presents a challenge in determining the survival of helimagnetic orders at such elevated pressures. However, a pivotal clue emerges from the SHG results at 7 GPa, as depicted in Figure 2b. A distinct six-petal pattern is evident for selected temperature points, including 15, 90, and 150 K. This pattern markedly differs from the previously observed distorted

"8" pattern characterizing the helimagnetic state. Moreover, at this pressure, the magnon signal as shown in Figure 2e remains observable. Hence, this transition indicates the vanishing or complete suppression of the helimagnetic state at approximately 7 GPa, with a subsequent transformation into a collinear antiferromagnetic. The same experimental phenomenon can be observed at 8.6 GPa (Figure 2b). This further confirms a phase transition from the helimagnetic to the collinear antiferromagnetic states at high pressures. At 6.5 GPa, a notable shift is observed in the maximum-to-minimum intensity ratio of the SHG signal, reducing significantly to 4:1. It is notably lower than the value observed at 2.7 GPa. Furthermore, the SHG pattern at 15 K deviates from its previously distorted '8' pattern, exhibiting a combination of '8' and six-petal patterns. One possible explanation could be the coexistence of AFM and helimagnetic phases due to uneven pressure distribution around the phase transition point. Alternatively, the deviation may originate from an intermediate state between the helimagnetic and AFM states, which will be discussed later. These alterations strongly indicate the transition from a helimagnetic to an antiferromagnetic state has occurred at this critical pressure.

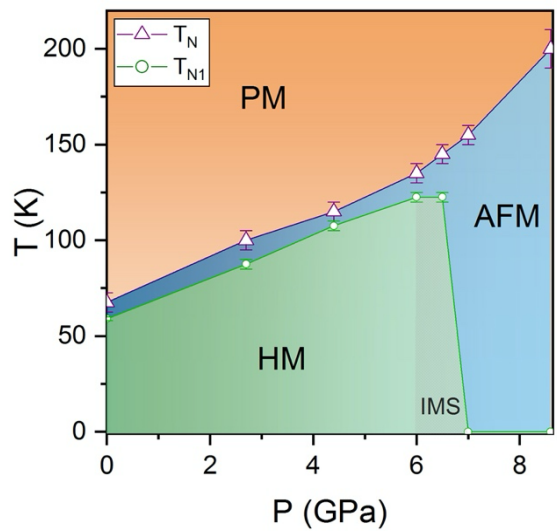

Figure 4 The magnetic phase diagram of the NiI$_2$ flake as a function of pressure and temperature. $T_N$ and $T_{N1}$ are obtained from Raman and SHG measurements, respectively. Here, PM represents the paramagnetic state.

As the pressure surpasses 10 GPa, the magnon signal becomes undetectable (Figure S2), possibly due to a deteriorating signal-to-noise ratio. Notably, Mössbauer spectroscopy, as reported in reference [31], has demonstrated the influence of different pressures on magnetic properties. Under 19 GPa, the Néel temperature has been observed to rise to 310 K. At 11 GPa, our SHG patterns, as depicted in Figure S5, exhibit a consistent six-petal pattern across the temperature range of 20-100 K, indicating the likely presence of a collinear antiferromagnetic state. In Figure 4, we compile a comprehensive phase diagram outlining the transition from the helimagnetic state to the antiferromagnetic state under varying pressures. Below 6 GPa, the transition temperature for the helimagnetic state (open green circles) exhibits a gradual ascent to 122.5 K with increasing pressure before transforming into an antiferromagnetic state at around 7 GPa. This transition is accompanied by an intermediate state (IMS in Figure 4) between two pressures. Furthermore, the antiferromagnetic state (purple triangles) endures at 190 K under 8.6 GPa pressure from our Raman findings. From Ref. [31], the actual transition from antiferromagnetic to paramagnetic state can increase to 216 K at around 8.3 GPa. Importantly, upon releasing the pressure to 2 GPa, the reappearance of the distorted "8" pattern below 81 K (as shown in Figure S6) underscores the pressure-controlled transition between the helimagnetic and antiferromagnetic states. This novel capability to modulate the helimagnetic state transition via pressure holds the promise of advancing innovative functional devices driven by external forces.

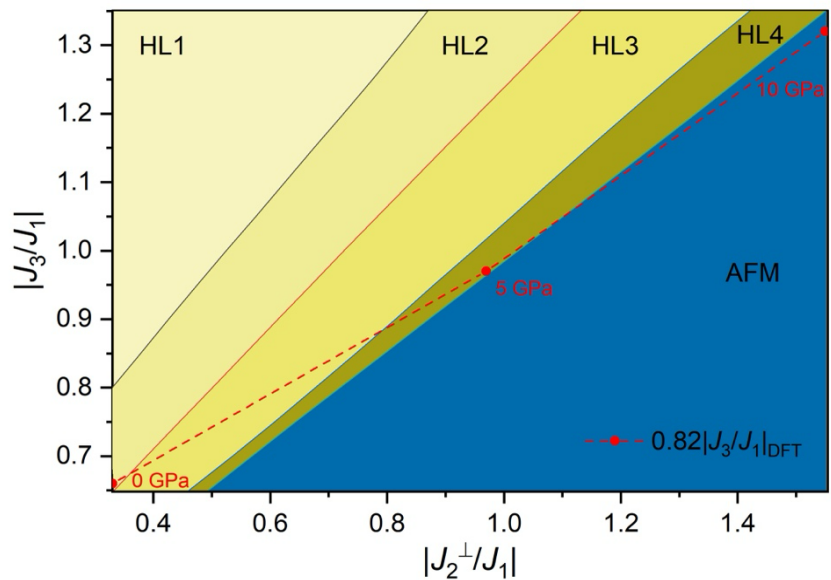

Figure 5: Phase diagram encompassing all the magnetic states (HL1-4 and AFM as described in Figure S7) with differing intralayer and interlayer exchange interactions. HL2 is identified as the helimagnetic ground state for NI$_2$ at low temperature. The red data points represent the exchange interaction under pressure, sourced from DFT calculations in Ref. [1]

Monte Carlo (MC) simulations were conducted to investigate the response of the magnetic properties of NiI$_2$ to external pressure. Our simulation incorporates consideration of the intralayer nearest-neighbor interaction ($J_1$), the intralayer next-next-nearest neighbor interaction ($J_3$), and interlayer next-nearest neighbor interaction ($J_2^\perp$) as defined in Ref. [1]. Other small exchange interactions estimated in DFT are also added and fixed during our simulation. Detailed information regarding the simulation methodology can be found in the Method section. At ambient pressure, the helimagnetic ground state, as indicated by the HL2 (Figure S7), is attained due to the competition between the intralayer FM interaction $J_1$ and the AFM interaction $J_3$ [1]. In this state, spins rotate within the canted $Ni_2I_2$ plane with a propagation vector Q ≈ (1/7, 0, 3/2), as depicted in Figure 1(b) and Figure S7. The spin configuration of the helimagnetic states is sensitive to the competition between $J_1$, $J_3$, and the intralayer interactions $J_2^\perp$. Varying the interlayer and intralayer interactions has yielded other helimagnetic states, denoted as HL1, HL3, and HL4 (Figure S7), with corresponding propagation vectors (1/14, 1/14, 1/2), (1/14, 6/7, 0.5), and (0, 13/14, 0.5), respectively. Remarkably, the AFM (Figure S7) state can emerge as the dominant magnetic ground state when the interlayer interaction is strengthened, as indicated in the deep blue region of Figure 5. It differs from the MC simulation in Ref. [1] in which only the helimagnetic states are revealed under high pressure. To incorporate the DFT results of the magnetic interactions [1] into our phase diagram, we scale the DFT values of $J_3$ by a factor $0.82J_3$ to align with our high-pressure experimental results of the magnetic transition from the helimagnetic states to the AFM state. The DFT simulation may overestimate $J_3$, regarding to the fact that the MC phase boundary is sensitive to the magnetic interaction parameters. In Figure 5, the DFT results with the scale ($0.82J_3$) indicate that the exchange interaction parameter at ambient pressure resides within the HL2 region. However, at pressures of 5 and 10 GPa, the results shift to the HL4 and AFM regions, respectively. This suggests that pressure induces concurrent increases in both the intralayer interaction $J_3$ and the interlayer interaction $J_2^\perp$, thereby causing a transition in the magnetic ground state from HL2 to eventual AFM under high-pressure conditions. This observation aligns with experimental findings, which note a transition to the AFM state above 7 GPa hydrostatic pressure. Of significant note, if the interlayer interaction remains constant and only the intralayer interaction $J_3$ is intensified, the magnetic state transitions exclusively from HL2 to HL1, without the potential to reach the AFM state. Consequently, we discern the substantial impact of the second-nearest neighbor interlayer interaction $J_2^\perp$, which competes with intralayer frustration and favors the AFM state. Additionally, MC simulations also indicate that under pressure, the sample undergoes transitions through the HL3 and HL4 states before ultimately reaching the AFM state. This observation suggests that these states with varying wave vectors (q) may serve as intermediate states, a notion supported by the SHG observation around 6.5 GPa, as depicted in Figure 3(b).

**Conclusion**

In this study, we explored the helimagnetic state of $NiI_2$ flakes when subjected to pressure using in situ high-pressure SHG and Raman spectroscopies. Our findings reveal a significant magnetic phase transition occurring at approximately 7 GPa, transitioning from a helimagnetic to antiferromagnetic state. Notably, we observed a marked increase in the magnetic transition temperature for both states. Through our MC simulation and reported DFT results, we observed a rapid strengthening of the interlayer second nearest-neighbor interaction under pressure, surpassing the intralayer exchange interaction. This change in interaction dynamics is pivotal, as it drives the transition from a helimagnetic to an antiferromagnetic state.

**Author contributions**

J. D. and J. M. conceived the project. J. D., Q. L., and W. S. designed and performed the experiments. X. Z., X. X., and L. W. provided and characterized the samples. K. X. and D. X. provided SHG calculations and fitting. B. X., Y. G., X. Z., and J. M. provided the theoretical support. All authors discussed the results and co-wrote the paper.

**Acknowledgment**

We would like to thank Prof. Haizhou Lu from SUSTech for helpful discussions. J.W.M. acknowledges the support from the

National Key Research and Development Program of China (Grant No. 2021YFA1400400), Shenzhen Fundamental Research Program (Grant No. JCYJ20220818100405013), the Guangdong Innovative and Entrepreneurial Research Team Program (Grants No. 2017ZT07C062), Shenzhen Key Laboratory of Advanced Quantum Functional Materials and Devices (Grant No. ZDSYS20190902092905285), Guangdong Basic and Applied Basic Research Foundation (Grant No. 2020B1515120100). J.F.D. acknowledges the support from the National Natural Science Foundation of China (11974159) and the Guangdong Natural Science Foundation (2021A1515012316). The theoretical part was supported by the National Natural Science Foundation of China (11974197 and 51920105002), Guangdong Innovative and Entrepreneurial Research Team Program (No. 2017ZT07C341).

**Conflict of Interest**

The authors declare no conflict of interest.

**Methods**

**Synthesis of $NiI_2$ single crystals**: The high-quality $NiI_2$ single crystals studied in this paper were prepared by the chemical vapor transport method. The Ni powder (99.999%) and the transport agent $I_2$ pieces were placed in an alumina crucible with a molar ratio of Ni : $I_2$ = 1 : 1.5. After evacuated and sealed into a quartz tube, the crucible was placed into a single-zone tube furnace with the source temperature of 700°C keeping for 7 days. A large number of millimeter-sized crystals were collected at the end of the quartz tube.

**SHG measurement**: The SHG measurements were conducted in a back-reflection geometry. The fundamental wave was generated from a Ti-Sapphire oscillator with an 80 MHz repetition rate and 150 fs pulse width. A wavelength of 880 nm was chosen for temperature-dependent SHG intensity and angular-resolved SHG measurements. The laser pulses were focused to a spot of around 2 $\mu m$ on the sample at normal incidence using a 50× objective. The reflected SHG signal was collected by the same objective and detected by a spectrometer equipped with a thermoelectric-cooled charged-coupled device (CCD). A shortpass filter with a central wavelength of 650 nm was used to block the fundamental wave. For angular-resolved SHG measurements, a 1/2-wave plate with a wavelength range of 310-1100 nm is used to adjust the polarization of the fundamental wave. The reflected SHG signal passed through the same 1/2 waveplate and was divided into parallel and perpendicular components by a displacer. By rotating the fast axis of the 1/2 waveplate, the angular-resolved SHG in two orthogonal directions are obtained.

**Raman measurement**: Raman spectra were performed using a homemade micro-Raman system in the back-scattering geometry. A He-Ne laser at 633 nm was employed to excite $NiI_2$ samples, and the laser power was kept below 200 μW to avoid sample heating. The laser beam passed through three ultra-narrow bandpass filters, and was focused onto the $NiI_2$ samples by a 50x objective. The back-scattered light was collected by the same objective, passed through three narrow notch filters to suppress the Rayleigh scattering, and then was focused on the entrance slit of a spectrometer with 1800 grooves $mm^{-1}$ diffraction grating and a nitrogen-cooled charge-coupled device (CCD).

**Theoretical calculations**:
We conduct Monte Carlo (MC) simulations on an $14 \times 14 \times 12$ three-dimensional triangular lattice with ABC stacking. The model Hamiltonian encompasses both isotropic and anisotropic exchange interactions. In our simulations, the intralayer nearest-neighbor interaction, denoted by $J_1$, is set to 1, serving as the energy unit. The intralayer nearest-neighbor anisotropic exchanges are set as follows: $J_{xx} = 0.1$, $J_{yy} = -0.12$, $J_{zz} = 0.03$, and $J_{yz} = 0.05$. Additionally, we consider two parameters for the intralayer next-next-nearest neighbor interaction $J_3$ and interlayer next-nearest neighbor interaction $J_2^\perp$. Furthermore, we maintain the magnitude of interlayer next-next-nearest neighbor interaction $J_3^\perp$ fixed at $0.35 J_2^\perp$. The phase diagram is obtained through a three-stage process. Initially, we employ a combination of the parallel tempering method and heat-bath sampling algorithm. Forty replicas of the system are simulated over a temperature range from T = 0.01 and T = 4, following a geometric progression. During this stage, we conduct $2 \times 10^5$ MC sweeps and exchange replicas after 100 sweeps. In the second stage, all replicas are quenched to a lower temperature of $T = 0.001$, followed by $10^5$ heat-bath sweeps. Finally, we perform zero-temperature sampling, which involves aligning spins with the direction of their local effective field $H_{eff}$ and employing an over-relaxation process, i.e., mirroring the spins according to $H_{eff}$. The ground states are then determined based on the lowest energy configurations among the 40 replicas.

Correlated Metallic State of the van der Waals Insulator ${\mathrm{CrGeTe}}_{3}$. *Physical Review Letters* **127**, 217203 (2021). https://doi.org:10.1103/PhysRevLett.127.217203